\documentclass[prl,twocolumn,amssymb,amsmath,showpacs]{revtex4}

\usepackage{graphicx}
\usepackage{amssymb}
\usepackage{amsmath}
\usepackage{color}

\usepackage{palatino}
\usepackage[T1]{fontenc}

\definecolor{violet}{rgb}{0.7,0,0.7}

\def\url#1{\textcolor{blue}{\underline{#1}}}	

\definecolor{oneblue}{rgb}{0,0,0.65}
\def\b#1{\textcolor{oneblue}{#1}}

\begin{document}


\title{A framework to reconcile frequency scaling measurements, from
  intracellular recordings, local-field potentials, up to EEG and MEG
  signals}

\author{Claude Bedard, Jean-Marie Gomes, Thierry Bal and Alain
Destexhe \\ \ \\UNIC, CNRS, Gif sur Yvette, France.\\ \ \\
Submitted to the special issue on {\it Extracellular Space} (Edited
by Reinoud Maex).}

\date{\today}

\begin{abstract}

In this viewpoint article, we discuss the electric properties of the
medium around neurons, which are important to correctly interpret
extracellular potentials or electric field effects in neural tissue.
We focus on how these electric properties shape the frequency scaling
of brain signals at different scales, such as intracellular
recordings, the local field potential (LFP), the electroencephalogram
(EEG) or the magnetoencephalogram (MEG).  These signals display
frequency-scaling properties which are not consistent with resistive
media.  The medium appears to exert a frequency filtering scaling as
$1/\sqrt{f}$, which is the typical frequency scaling of ionic
diffusion.  Such a scaling was also found recently by impedance
measurements in physiological conditions.  Ionic diffusion appears to
be the only possible explanation to reconcile these measurements and
the frequency-scaling properties found in different brain signals.
However, other measurements suggest that the extracellular medium is
essentially resistive.  To resolve this discrepancy, we show new
evidence that metal-electrode measurements can be perturbed by shunt
currents going through the surface of the brain.  Such a shunt may
explain the contradictory measurements, and together with ionic
diffusion, provides a framework where all observations can be
reconciled.  Finally, we propose a method to perform measurements
avoiding shunting effects, thus enabling to test the predictions of
this framework.

\end{abstract}

\maketitle



\vspace{5mm}


\section{Introduction}

The electric nature of the extracellular medium around neurons is of
high importance to correctly interpret the extracellular potentials,
such as the local field potential (LFP), as well as more remote
potentials, such as the electro-encephalogram (EEG).  This electric
nature can be captured by appropriate measurements of the
extracellular impedance.  However, the measurements available today,
and their interpretation, are contradictory.  While some measurements
suggest that the extracellular medium is essentially
resistive~\cite{Ranck,Logothetis}, other
measurements~\cite{Schwan,Gabriel,Nelson2013,gomes2016} suggest that
the medium is non-resistive, and strongly frequency dependent.  There
is presently no consensus on this electric nature.

On a theoretical point of view, in the neuronal cable theory initially
developed by Rall~\cite{Rall1962,Rall1995}, the extracellular medium
was assumed to have zero resistance, and the neuron was thus
considered as embedded in a supraconductive medium.  Later
formulations~\cite{Plonsey,Tuckwell} included a resistance to represent the
medium, but it was always assumed that this resistance is much smaller
than that of the membrane.  The whole development of cable theory was
made under this assumption, and to include non-resistive media in
cable equations requires to re-derive the equations from first
principles.  This was done recently, leading to the ``generalized
cable theory''~\cite{BedDes2013,BedDes2014}, that provided a cable
theory valid for arbitrarily complex extracellular media (and includes
Rall's cable theory as a particular case).  It was found that the
extracellular impedance appears in the length constant of the neuron,
and thus the nature of the medium potentially can influence the
integrative properties of neurons~\cite{BedDes2013}.

Modeling complex extracellular media started with an initial model
that only considered the impedance inhomogeneities (such as fluids and
membranes), and it was found that such inhomogeneous structure can
lead to strong frequency filtering
effects~\cite{BedDes2004,BedDes2009,ChapLFP2009}.  It was later shown,
using a mean-field formalism, that various physical processes such as
polarization~\cite{BedDes2009,BedDes2006b} or ionic
diffusion~\cite{BedDes2009,BedDes2011a} can similarly cause frequency
filtering, and thus influence the genesis of the LFP.  {It was
  shown that a medium with polarization is equivalent to a
  resistance-capacitance circuit~\cite{BedDes2006b}, thus exerting
  strong low-pass filtering on extracellular potentials.  Similarly, a
  medium with diffusive properties will also exert a low-pass
  filtering.  None of such filtering is present with a resistive
  medium.}

The nature of the medium can also influence the estimation of neuronal
sources from extracellular recordings, thus affecting methods such as
the Current Source Density (CSD) analysis~\cite{BedDes2011a}.  Like
cable equations, the CSD method assumes a resistive
medium~\cite{Mitzdorf85}, and is not valid for more complex
extracellular properties.  The CSD method was generalized by
rederiving the equations from first principles, yielding a generalized
CSD which includes the classic CSD as a particular case, and which can
estimate neuronal sources within a non-resistive extracellular
medium~\cite{BedDes2011a}.  Here again, it was found that the nature
of the medium has potentially large influences on the CSD estimates of
neuronal sources.

This underlines the importance of having a precise estimate of the
impedance of the extracellular medium.  In the present paper, we
review a number of measurements at different scales, from
intracellular, to LFP, and up to large scale such as the EEG.  We also
provide new analyses of experimental results and propose a framework
where all contradictory data can be explained.


\section{Results}

We first review evidence that different brain signals, such as
intracellular recordings, the local field potential (LFP), the
electroencephalogram (EEG) or the magnetoencephalogram (MEG), all
display properties that are not consistent with resistive media. 
We next show results from extracellular metal-electrode
measurements that suggest a way to resolve discrepancies between
different measurements in the literature.

\subsection{Frequency scaling of different brain signals}

\subsubsection{$1/f$ scaling of EEG and LFP}

\begin{figure}[t!]
\begin{center}
 \includegraphics[width=\columnwidth]{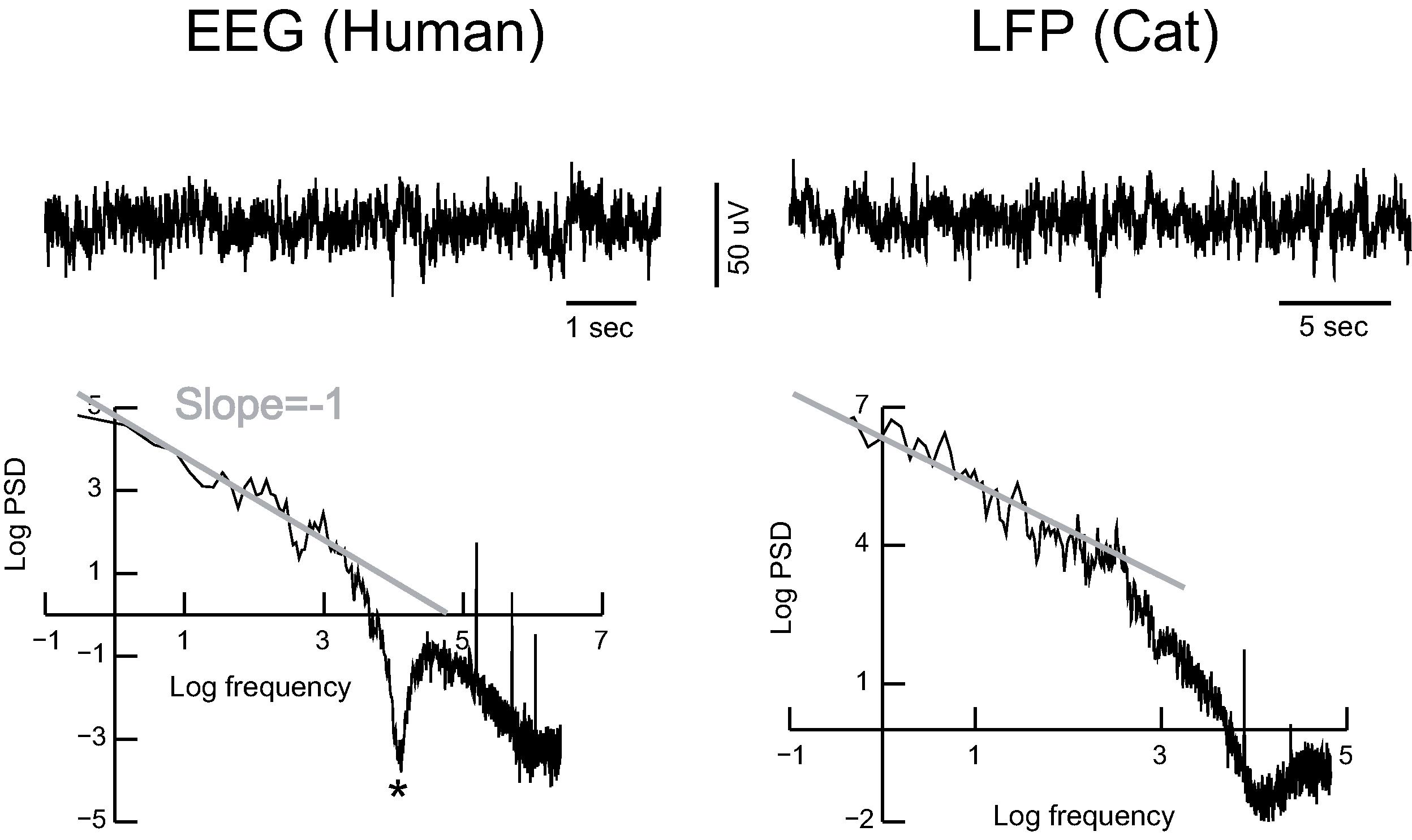}
\end{center}

\caption{$1/f$ Frequency scaling of EEG and LFP signals in awake
  subjects.  Top: human EEG recording (left, vertex EEG) and LFP
  recording from cat parietal cortex (right) in awake and attentive
  subjects (desynchronized EEG). Bottom: the corresponding power
  spectra display approximate $1/f$ scaling at low frequencies.  The
  straight lines (gray) indicate a slope of -1 (log-log
  representation).  The signals were not filtered, except for a notch
  filter at 60 Hz (*) for the EEG.  {All power spectra were
    computed using the Fast Fourier Transform (FFT) algorithm, and
    were not normalized.}}

 \label{eeg}
\end{figure}

EEG and LFP signals can show $1/f$ frequency scaling properties at low
frequencies ($<$10~Hz), as illustrated in Fig.~\ref{eeg}.  This was
shown by a number of studies
\cite{Pritchard92,Novikov97,Bha2001,BedDes2006a,Dehgha2010a}.  It is
important to note here that such $1/f$ scaling depends on brain state
and is seen in awake subjects with strictly desynchronized EEG.  In
other brain states, the frequency scaling may be different, for
example during anesthesia the EEG scales as $1/f^2$
\cite{Milstein2008}.  The frequency scaling illustrated here in the
human EEG and in the LFP recorded in cat parietal cortex was done in
subjects that were awake and attentive.  The exact value of the
exponent also varies as a function of the brain region considered, as
also shown previously \cite{Dehgha2010a}.

\subsubsection{LFP-unit measurements} \label{lfp}

\begin{figure}[t!]
 \includegraphics[width=\columnwidth]{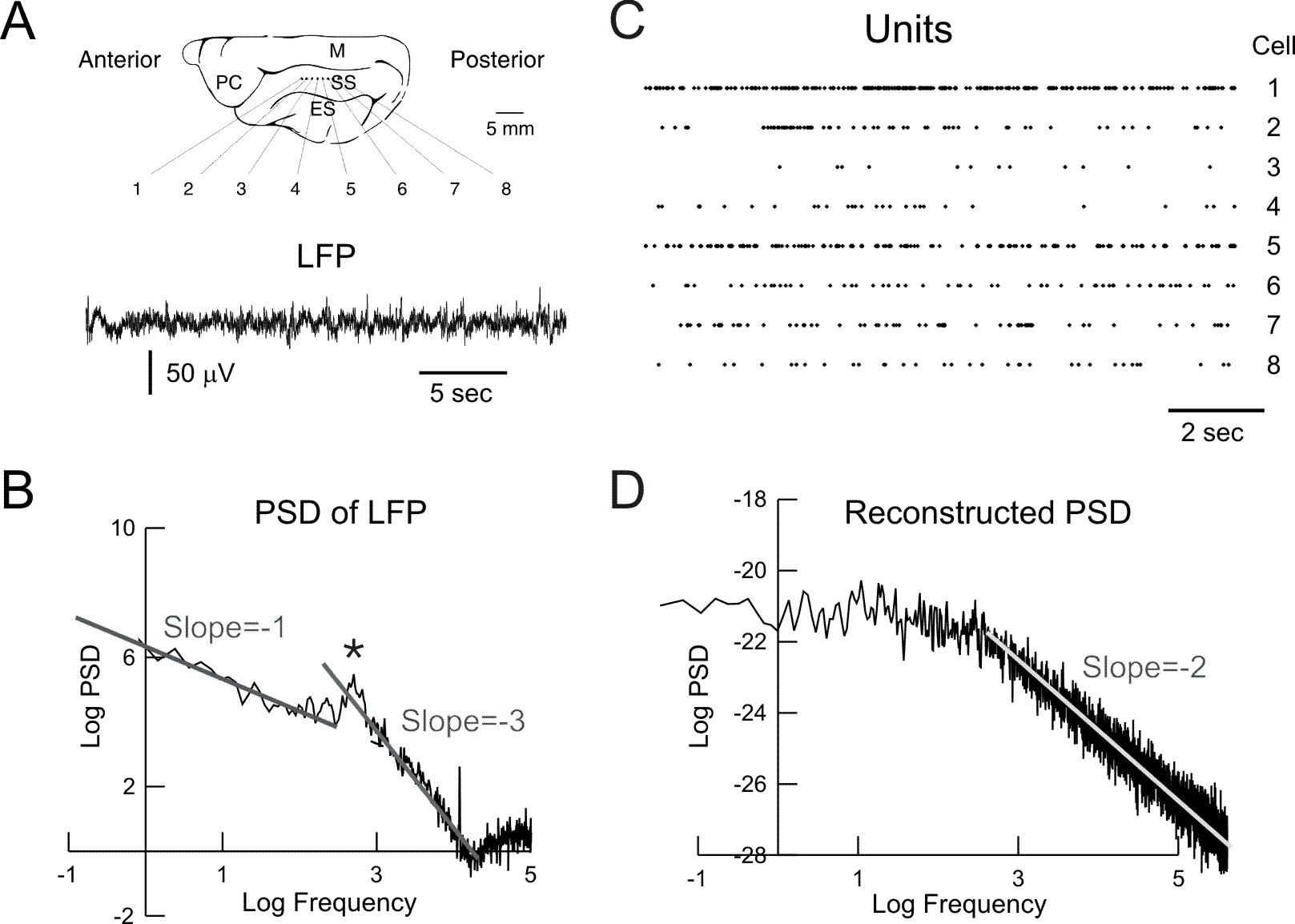}

\caption{Relationship between unit activity and LFP power spectra.  A.
  LFP recording in awake cat parietal cortex (same recording as in
  Fig.~\ref{eeg}, right). The top scheme shows the location of {the}
  electrodes in parietal cortex.  B. Power spectral density (PSD) of
  the LFP, showing that low frequencies scale as $1/f$ (gray line,
  slope=-1), and $1/f^3$ at higher frequencies (gray line; slope=-3).
  C.  Unit activity from the same experiment, recorded with a system
  of 8 tungsten electrodes (schematized in A). D.  Attempt to
  reconstruct the LFP signal from the unit activity. The low frequency
  end of the PSD was constant (zero slope), while the high-frequency
  end scaled as $1/f^2$ (gray line, slope=-2). An exponent of -1 is
  missing to reproduce the LFP scaling, which could be the sign that
  the current sources are subject to of an $1/f$ filter (modified from 
  Bedard et al \cite{BedDes2006a}.}

 \label{lfpunits}
\end{figure}

Different mechanisms were proposed to explain the origin of such
``$1/f$ noise'' in the brain. $1/f$ spectra can result from
self-organized critical states \cite{Jensen98}, suggesting that
neuronal activity may be working according to such states
\cite{Beggs2003}, but this subject is controversial
\cite{BedDes2006a,Dehgha2012}. The morphology of the neuron may also
be responsible for filtering in the $1/f$ to $1/f^2$ range
\cite{Pettersen2008}, but this scaling applies to high frequencies and
cannot explain the $1/f$ scaling at low frequencies.  Finally, the
$1/f$ scaling may be due to filtering properties of the currents
through extracellular media \cite{BedDes2006a}. This conclusion was
reached by noting that the global activity reconstructed from
multisite unit recordings scales identically as the LFP if a ``1/f
filter'' is assumed, and without the need to assume self-organized
critical states in neural activity (Fig.~\ref{lfpunits}). However, the
latter study made the point that $1/f$ filtering may be necessary to
explain the experimental results, but no mechanism was provided.  We
will show below that ionic diffusion can explain such a $1/f$ filter.

\subsubsection{Modeling the $1/f$ scaling of LFPs}

\begin{figure}[t!]
 \includegraphics[width=\columnwidth]{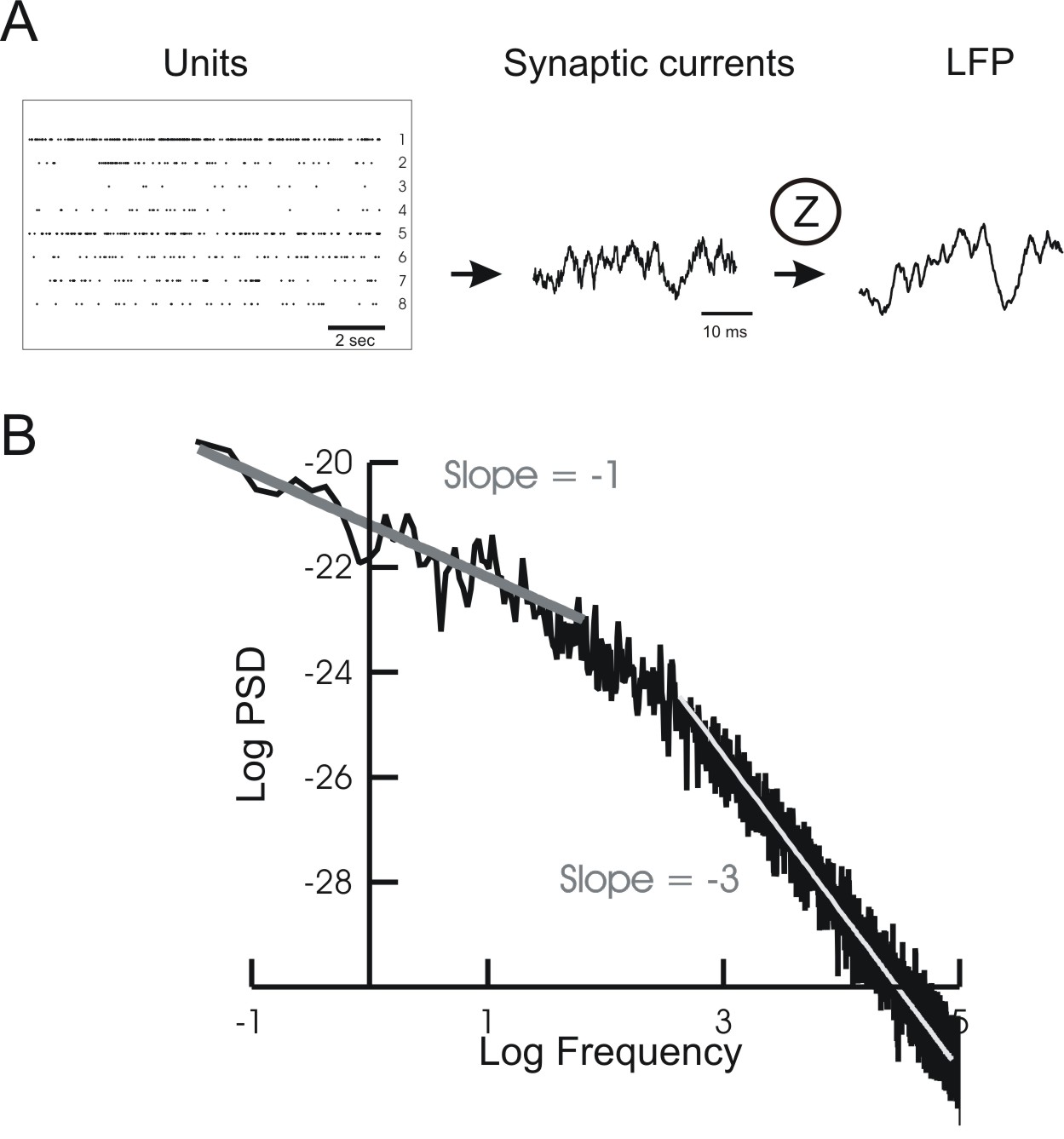}

 \caption{Reconstruction of LFP power spectra from unit activity using
   ionic diffusion. A. Scheme of the reconstruction.  The unit
   activity is used to generate a synaptic current.  The current is
   used in a model of LFP that uses a diffusion impedance ($Z$,
   varying as $1/\sqrt(\omega)$).  B. PSD of the modeled LFP, which
   qualitatively displays the same frequency scaling as the real LFP
   (compare with Fig.~\ref{lfpunits}B; modified from Bedard and
   Destexhe, \cite{BedDes2009}).}

 \label{recon}
\end{figure}

$1/f$ scaling in power spectra is not easy to explain, because it
predicts a filter in $1/\sqrt{f}$.  Classic filters such as a
capacitive filter, or an RC-circuit such as in neuronal membranes,
would predict $1/f^2$ filtering in power spectra.  It was shown that
ionic diffusion can generate frequency scaling as $1/\sqrt{f}$
\cite{BedDes2009,Diard1999,Tay1995}.  Using a macroscopic modeling 
approach based on a mean-field formulation of Maxwell 
equations\cite{BedDes2009,BedDes2011a}, it was shown that ionic 
diffusion can give rise to $1/f$
frequency scaling at low frequencies (Fig.~\ref{recon}).  This scaling
arises because the ionic diffusion in the extracellular medium and
around the current sources is responsible for a ``diffusion
impedance'' scaling as $1/\sqrt{f}$, which gives $1/f$ in the power
spectrum.  For high frequencies, the natural $1/f^2$ scaling of
current sources (which are mostly exponential) is also subject to the
same filter, which gives the observed $1/f^3$ scaling.  With ionic
diffusion, one can qualitatively reconstruct the frequency scaling of
LFPs from the unit activity, and thus, ionic diffusion appears as a
physically plausible explanation for the observed scaling.

\subsubsection{Intracellular-LFP measurements} \label{transfer}

\begin{figure}[t!]
 \includegraphics[width=\columnwidth]{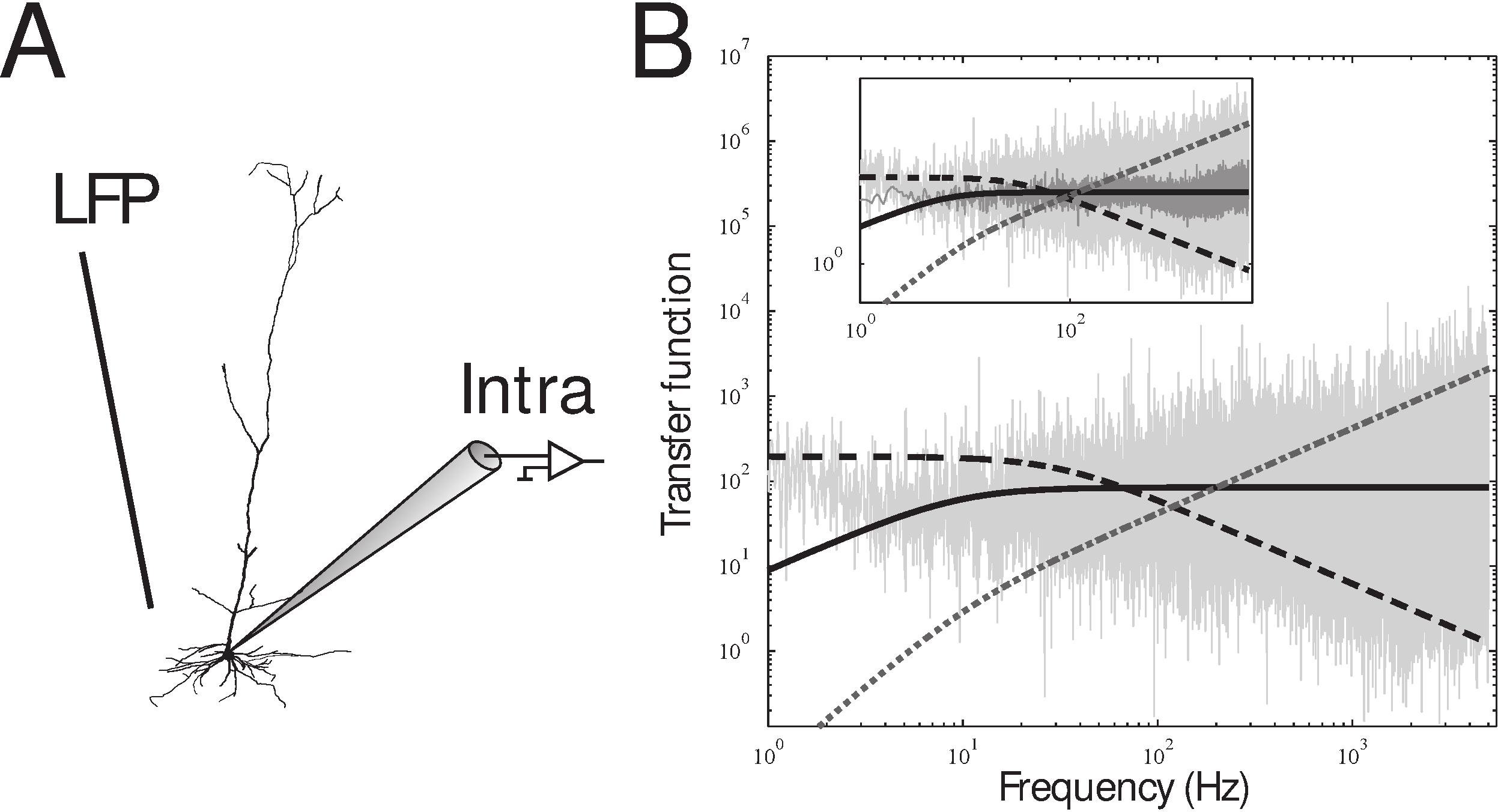}

 \caption{Transfer function between intracellular and extracellular
   potentials {\it in vivo}.  A. Scheme of the recording, where
   intracellular and LFP measurements are made within a close
   vicinity.  B. Transfer function spectrum computed from a cell
   recorded in rat barrel cortex {\it in vivo} during desynchronized
   EEG states (gray spectrum).  The black lines show different
   transfer functions calculated from a ball-and-stick model
   surrounded by media with different impedances, resistive (dashed),
   diffusive (solid), and capacitive (dotted).  Modified from Bedard
   et al.,\cite{BedDes2010c}.}

 \label{transf}
\end{figure}

To further probe the LFP signal, we used simultaneous intracellular
and LFP measurements, as schematized in Fig.~\ref{transf}A.  In
particular, it is interesting to focus on the transfer function
between intracellular and extracellular signals.  This transfer
function was evaluated from simultaneous intracellular and LFP
measurements in rat barrel cortex {\it in vivo} \cite{BedDes2010c}
and is represented in Fig.~\ref{transf}B (gray).  The interest of this
approach is that when relating intracellular and extracellular
voltages, the impedance of the extracellular medium naturally appears,
so matching different models to the measured transfer function allows
one to estimate the extracellular impedance.  This estimate is
indirect, however, because this model is valid only for brain states
where neuronal activities are perfectly decorrelated.  This is
why this estimate must be performed in desynchronized-EEG brain
states, as done in Bedard et al \cite{BedDes2010c}.  Different hypotheses about
the extracellular impedance are shown in Fig.~\ref{transf} (black
curves).  Neither resistive nor capacitive media provided acceptable
fits, while the best match was obtained for an impedance scaling as
$1/\sqrt{f}$, compatible with ionic diffusion.  Thus, similar to the
spectral analysis of LFPs, the LFP-intracellular relation is also
compatible with an electrical impedance with strong ionic diffusion
effects.

\subsubsection{EEG-MEG measurements} \label{meg}

\begin{figure}[t!]
 \centering
 \includegraphics[width=\columnwidth]{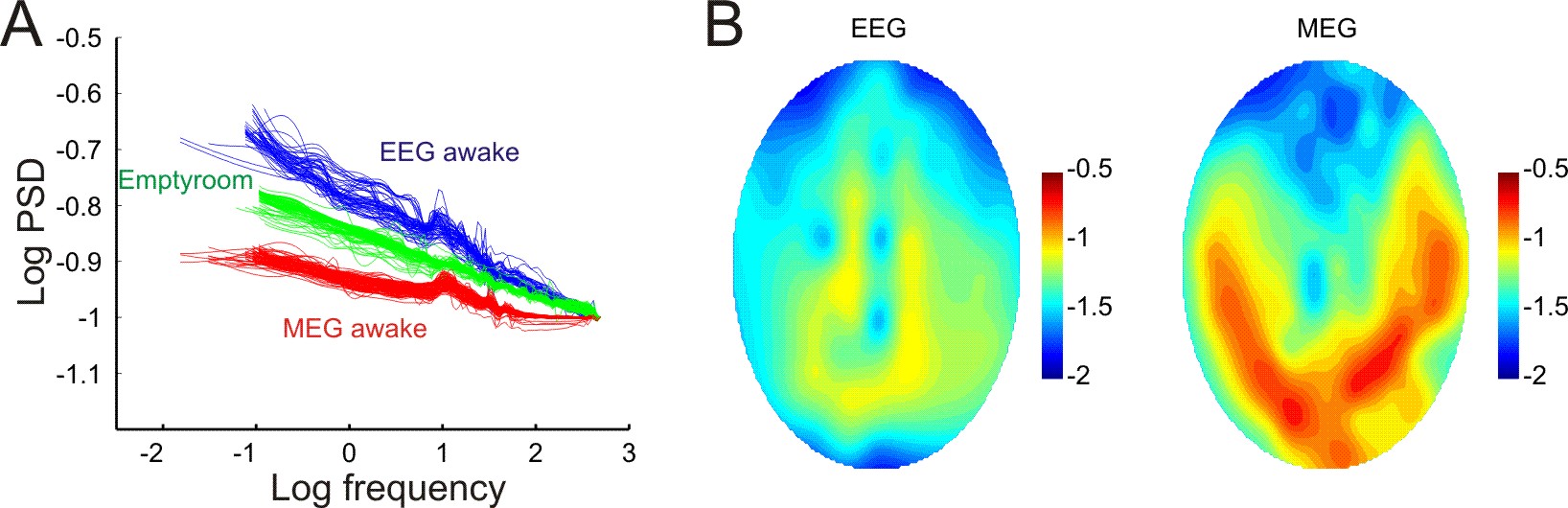}

 \caption{Different frequency scaling of EEG and MEG signals.  A.
   Frequency spectra of simultaneously-recorded EEG (blue) and MEG
   (red) signals from an awake subject with desynchronized EEG. The
   ``empty-room'' MEG signal (green) is also shown for comparison.  B.
   {Left:}
   Distribution of low-frequency scaling exponents over the scalp for
   the EEG, showing that the low-frequency scaling exponent is
   comprised mostly between 1 and 2.  {Right:} Same representation
   for MEG signals.  In this case, the exponent is lower, and
   generally {smaller} than 1.  Modified from Dehghani et 
   al.~\cite{Dehgha2010a}.}

 \label{eegmeg}
\end{figure}

Another type of signal that can be used to infer the nature of
extracellular space is the magnetic field generated by neuronal
activities.  In particular, using simultaneously-recorded EEG and MEG
signals, it is possible to relate their frequency scaling properties.
Theoretical work shows that, if the extracellular medium is resistive,
the scaling of EEG and MEG signals at low frequencies should be the
same \cite{Dehgha2010a}.  Similarly to above for LFP and
intracellular signals, this relation is only valid assuming that the
synaptic inputs are uncorrelated, so it should be evaluated in brain
states as desynchronized as possible.  The measurement of the
frequency scaling in awake human subjects (with desynchronized EEG)
showed that the frequency scaling is generally not the same between
EEG and MEG signals (Fig.~\ref{eegmeg}).  The difference is evident by
visual inspection of superimposed spectra (Fig.~\ref{eegmeg}A), and this
difference is confirmed by the distribution of scaling exponents in
different brain regions (Fig.~\ref{eegmeg}, B).  A detailed analysis
showed that when the exponents were similar, the signal to noise
ratio was very low, and that this difference is significant 
\cite{Dehgha2010a}.  Thus, the relation between EEG and MEG signals suggests
that the extracellular medium is not resistive, although this analysis
does not say more about which type of medium is the most likely.

\subsubsection{Intracellular measurements} \label{intra}

\begin{figure}[t!]
 \centering
 \includegraphics[width=\columnwidth]{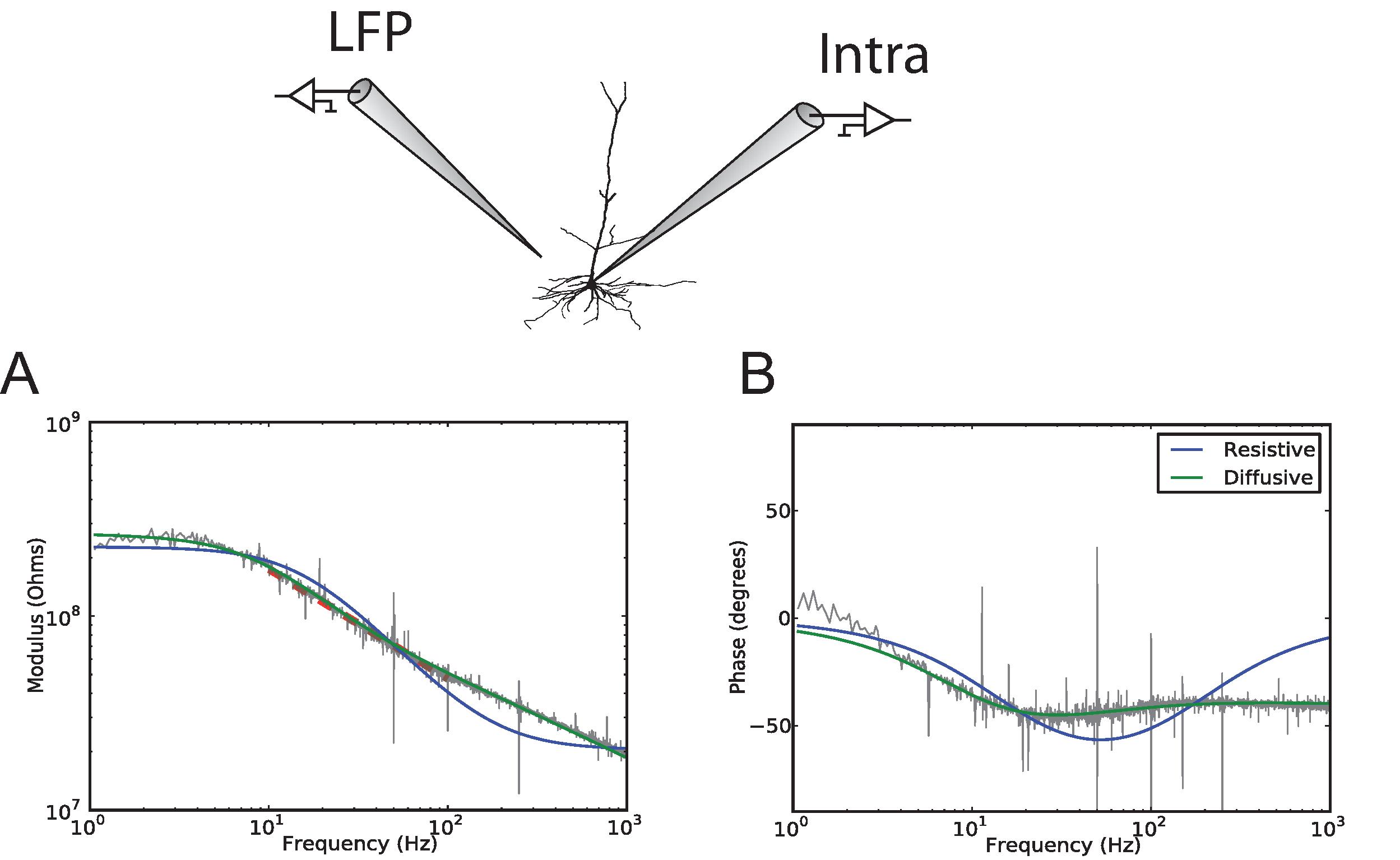}

 \caption{Natural impedance measurement {\it in vitro}.  Top:
   experimental setup, where a subthreshold white-noise current was
   injected in a cell {in cortical slices}, together with an
   extracellular recording in the
   vicinity using a second micropipette located about 20 $\mu$m away
   from the soma of the patched cell. The modulus (A) and phase (B) of
   the measured impedance are shown as a function of frequency.  The
   colored curves show the best fits using a resistive model (blue)
   and a diffusive model (green).  Modified from Gomes et 
   al.~\cite{gomes2016}.}

 \label{imped}
\end{figure}

Finally, the measurement of the extracellular impedance can be done
directly using two micropipettes, as illustrated in Fig.~\ref{imped}
(top scheme).  Here, the intracellular recording was performed in
reference to a nearby micropipette in the extracellular medium, and a
subthreshold white noise current input was injected into the cell.
The relationship between the injected current, and the difference
between intracellular and extracellular voltages, gives a direct
access to the extracellular impedance.  This measurement is done here
in natural conditions, because no current is injected in the
extracellular space, and the amount of current is also much smaller
and is within the physiological range.  This is different from the
classic metal-electrode measurements of impedance, which must use
artificially high currents, and also involve a complex interface
between the metal and the living medium.  This ``natural'' impedance
measurement \cite{gomes2016} is therefore more accurate and more
physiological because the current source in the medium is the neuron
itself, using all the natural biochemical and biophysical mechanisms
of how cells interact with the surrounding medium.

It is important to note that this measurement is very different from
that of Fig.~\ref{transf}, although both involve simultaneous
intracellular and extracellular recordings.  In the case of
Fig.~\ref{transf}, the ongoing activity is analyzed, and different
guesses for the extracellular impedance are compared to the measured
transfer function, whereas in the present case, the impedance is
directly measured by controlling the injected current.

An example of the measured impedance for a representative cell is
shown in Fig.~\ref{imped}.  The measurement of the impedance modulus
amplitude (Fig.~\ref{imped}A) and phase (Fig.~\ref{imped}B) show a
frequency profile that significantly departs from that predicted by a
resistive impedance (blue curves).  On the other hand, a diffusion
impedance accurately predicts the measured frequency profile
(Fig.~\ref{imped}A-B, green curves).  The same result was also
obtained by white-noise current injection {\it in vivo}, or by
injection of sinusoidal currents {\it in vitro} \cite{gomes2016}.

\subsection{Metal-electrode measurements}

The results reviewed in the previous section show that the frequency
scaling of different brain signals, from microscopic to macroscopic
scales, all point to the fact that the medium is well described by a
diffusion impedance.  However, this result is not in agreement with
previous measurements using metal electrodes, suggesting a resistive
extracellular medium \cite{Logothetis}.  To further
investigate this issue, we have performed additional experiments.

\begin{figure}[t!]
 \centering
 \includegraphics[width=\columnwidth]{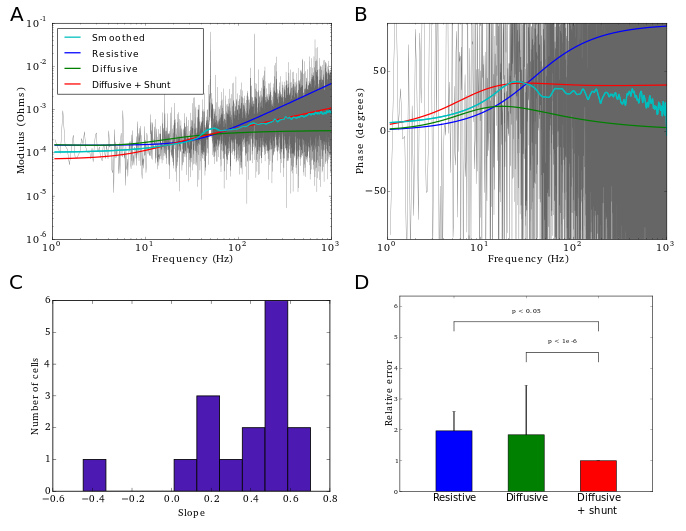}

 \caption{Transfer function between intracellular and extracellular
   potentials {\it in vitro}. A: Modulus of the transfer function
   $V_{LFP} / V_{intra}$ as a function of frequency (gray).  B. Same
   as in A, but for the phase.  In A and B, the colored curves are
   respectively: resistive model (blue), diffusive model (green) and a
   model including ionic diffusion and a possible shunt in the
   measurement (red).  C.  Distribution of the frequency scaling
   exponent found for different cells.  D. Relative error of different
   models with respect to the data.}

 \label{shunt}
\end{figure}

Using the same setup as schematized in Fig.~\ref{imped}, it is also
possible to measure the transfer function of the system, as
illustrated in Fig.~\ref{shunt}A-B.  In this case, we have fit the
measured function with the same models as before, a resistive and a
diffusive model, as shown by the blue and green curves in
Fig.~\ref{shunt}A-B, respectively. Similar to above, the diffusive
model provided a better fit of the transfer function, but the
difference was essentially due here to the phase of the transfer
function (whereas in Fig.~\ref{transf}, only the PSD was shown).

\begin{figure}[t!]
 \centering
 \includegraphics[width=\columnwidth]{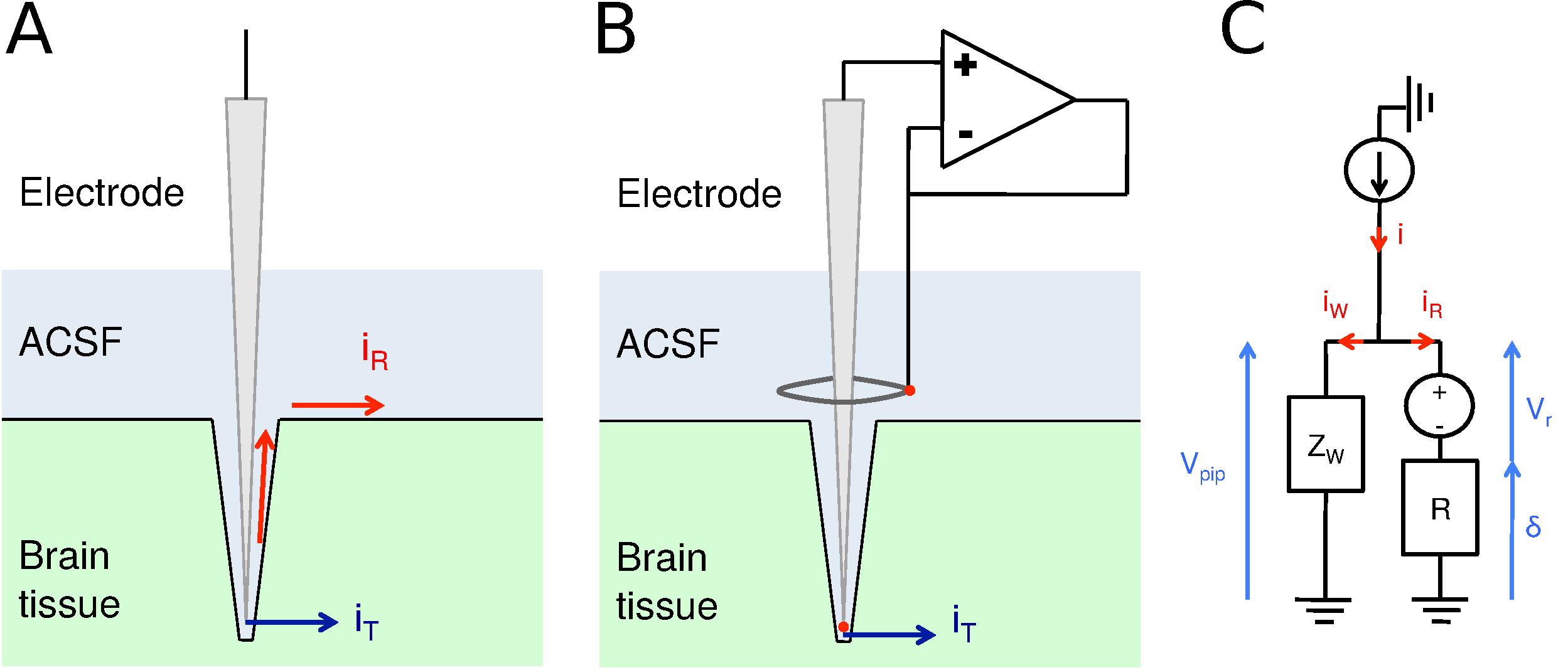}

 \caption{Experiment to test the presence of a shunt. A. When an
     electrode is inserted in brain tissue to inject a current, this
     current flows in the tissue (\b{$i_T$}) and in a nearly-resistive
     fluid layer (ACSF) flowing through the surface (\b{$i_R$}). B. Setup
     where a voltage follower imposes to the surface the same
     potential as that of the tip of the electrode (red dots).  This
     prevents currents from flowing in the fluid layer and forces the
     current to flow into the brain tissue. C. Equivalent circuit of
     this setup. The round shapes with an arrow and +/- signs are
     respectively current and voltage sources.}

 \label{anneau}
\end{figure}

To better explain these results, we made the following hypothesis, as
schematized in Fig.~\ref{anneau}A.  Metal electrodes, due to their
large diameter (microns), are necessarily surrounded by a thin layer
of cerebro-spinal fluid (CSF; also called {\it artificial CSF} or ACSF in 
the slice), and thus when injecting currents in a metal electrode, part 
of the current flows through the tissue, but another part of the 
current may also flow through the CSF (red arrow in Fig.~\ref{anneau}A), 
thereby creating a shunt.  Such a shunt will necessarily be resistive 
because the current flows only through the CSF liquid.  To test whether 
such a hypothesis is plausible, we included a resistive shunt in 
parallel to the diffusive impedance.  Such a ``diffusive + shunt'' 
model was able to better fit the measured transfer function (red 
curves in Fig.~\ref{shunt}).  In particular, the error was much smaller 
by using such a shunt (Fig.~\ref{shunt}D).

This measurement and fitting suggest that previous metal electrode
measurements may give the impression that the medium is resistive,
because part of the current goes through the CSF.  This situation was
examined in more detail in the Appendix, where we show that the
measured impedance will be a combination of the tissue impedance and
the impedance of the CSF, so all depends on the ratio of currents
that flow in each medium.

Finally, we would like to propose a way to experimentally avoid this
shunting effect through the CSF, as illustrated in
Fig.~\ref{anneau}B-C.  To avoid that current flows through the CSF,
one could use another electrode and a voltage-follower circuit to
clamp the voltage at the surface to the same value as that of the tip
of the electrode.  This additional electrode could take the form of an
ring around the metal electrode.  This way, because the surface and
the tip will be at the same voltage, there will be no current flow
through the CSF, and this should force the current to flow through
the extracellular medium.


\section{Discussion}

In this paper, we have reviewed different measurements, from
single neurons to large-scale recordings, which all converge to the
same conclusions: (1) the extracellular medium around neurons cannot be
considered as purely resistive; (2) all results can be explained 
assuming a frequency filter scaling as $1/\sqrt{f}$; (3) ionic diffusion 
appears as the physical process that explains most of these results. 
This includes the correlation between single-cells and LFPs, both at the
level of the transfer function \cite{BedDes2010c}, and direct
impedance measurements \cite{gomes2016}.  It also accounts for the
$1/f$ scaling of LFPs, and its relation with the unit activity 
\cite{BedDes2006a}, as well as for the relation between EEG and MEG
signals, that scale differently at low frequencies~\cite{Dehgha2010a}.

However, although these results cumulate into a quite strong evidence,
they do not constitute a proof that the medium is diffusive.  This is
in part because most of these results were obtained in
ongoing-activity conditions, where multiple sources were present in
the neuron, and were not controllable.  An exception is the impedance
measurement {\it in vitro} \cite{gomes2016}, where a single source was
present and controlled.  In this case, the current source was known,
as well as the intracellular and extracellular voltage, and their
modulus and phase relations showed particular frequency profiles, that
only ionic diffusion was able to capture.  It may be that taking into
account the dendritic filtering effect \cite{Pettersen2008} accounts
for parts of these results as well (T.\ Ness and G.\ Einevoll, private
communication).  However, it was found that the dendritic filtering
effect vanishes under {\it in vivo} conditions (see Fig.~2 in
\cite{BedDes2010c}), and the same measurements were also obtained {\it
  in vivo} \cite{gomes2016}, which suggests that dendritic filtering
is not a likely explanation for those results.  In addition, dendritic
filtering does not explain the low-frequency scaling, nor the
difference of scaling between EEG and MEG signals, and it was shown
explicitly~\cite{BedDes2010c} that it cannot account for the transfer
function measurements {\it in vivo}.

A main prediction from the impedance measurements is that the
extracellular impedance is of the same order -- or even larger -- than
that of the cell membrane, contrary to previous measurements.  This is
completely opposed to the classic view of a very low extracellular
impedance, which is usually neglected in the cable formalism for
modeling neurons \cite{Rall1962,Rall1995,Tuckwell}.  According to this
classic view, the extracellular medium is a by-pass, even considered
with zero resistance (supraconductive) in some cable formalisms .

In the present paper, we provided a first test of this prediction.  By
using experiments with two micropipettes~\cite{gomes2016}, one
intracellular (whole-cell) and one extracellular, we could evaluate
the transfer function {\it in vitro} (\b{$V_{LFP}/V_{intra}$}) (see
Fig.~\ref{shunt}).  This experiment shows that the difference of
impedance as estimated from these two electrodes is very small, which
confirms the prediction.


To provide a plausible way to resolve the discrepancy between these
experiments and previous measurements (e.g., \cite{Logothetis}), we
proposed a possible explanation based on a resistive current shunt via
the CSF on the surface of the brain (or ACSF in the slice), as
schematized in Fig.~\ref{anneau}A.  This resistive shunt, combined
with ionic diffusion, provides a better fit of the measured transfer
function (Fig.~\ref{shunt}), but the improvement of the fit is not by
itself a proof of the existence of such a shunt, so it remains a
prediction.  It does provide an explanation for why some of the
measurements of extracellular impedance concluded on a resistive
medium.  We hypothesize that, in these measurements, the part of the
current flowing through the CSF was large, so that the high impedance
of the extracellular medium was basically invisible.  In addition to
proposing this shunt hypothesis, we also suggested a method to
evaluate this effect experimentally (Fig.~\ref{anneau}).  We hope that
further experiments will use that method in order to clarify the issue
and explain the contradictory measurements.

Why is the extracellular medium characterized by a diffusive
impedance~?  There are currently two -- non exclusive -- possible
theoretical explanations.  The first possible explanation is that
ionic diffusion acts at the source of the current, in or near the
transmembrane ion channels.  It is well known that ionic diffusion is
central to establish and maintain the membrane potential~\cite{Hille},
and ionic diffusion is also necessarily implicated in re-equilibrating
the ionic concentrations, and maintaining the Debye layer in the
vicinity of the membrane.  It is thus possible that the visible
current source in the extracellular medium contains an important
contribution from ionic diffusion, which may explain why this
component is seen in the measurements.  This of course would consider
that the current flows in an essentially resistive extracellular
medium.

A second possible explanation does not postulate any special effect of
ionic diffusion at the source of the current, but how the current
flows in the extracellular medium.  The current flow is necessarily
associated with an electric field, and the field lines will depend on
the charge distribution in the cell, and the flow of charges will
follow these field lines.  However, the field lines will in general
not respect the complex shape of the interstitial space in the
extracellular medium.  Thus, the charges that follow the field lines
will necessarily meet obstacles (such as cell membranes or vessels),
and produce local concentration inhomogeneities.  Such concentration
gradients will implicate ionic diffusion.  Thus, ionic diffusion will
be the mechanism that will allow the charges to circumvent the
obstacles, and this may explain why the impedance is high, and why it
has a diffusive component.  {Future experiments should be designed to
  further test these possible mechanisms.}

{Finally,} it is important to mention that ionic diffusion combined
with a shunt, is so far the only coherent framework in which all the
experimental measurements find a possible explanation.  Other
hypotheses, such as the resistive medium or the dendritic filtering,
cannot explain some of the data.  This of course does not mean that
ionic diffusion is the correct framework, but we hope it will motivate
further experiments to clarify {the exact electrical nature of the
  extracellular medium.}

{In conclusion, we have shown that since the classic work on
  cable equations~\cite{Rall1962,Rall1995,Plonsey,Tuckwell} and CSD
  analysis~\cite{Mitzdorf85}, which all considered that the
  extracellular medium is resistive, there is quite substantial
  evidence for deviations from resistivity.  If the medium is non
  resistive, all the above formalisms are invalid and must be
  re-derived from first principles (Maxwell equations).  This was done
  for cable equations~\cite{BedDes2013} and CSD
  analysis~\cite{BedDes2011a}, which were generalized to be valid with
  arbitrarily complex extracellular media.  In the present paper, we
  review that indeed, there is quite strong evidence that the medium
  may be diffusive instead of resistive.  Thus, we conclude that
  experiments should now focus on evaluating the possible consequences
  of such non-resistivity on the integrative properties of single
  neurons, as well as on the genesis of extracellular potentials.}


\subsection*{Acknowledgments}

Research funded by the CNRS, the Paris-Saclay excellence network
(IDEX) and the European Community ({\it Human Brain Project}, 
H2020-720270).


\appendix

\section{Appendices}

\subsection{Impedance measurement in the presence of a shunt}
\label{appx}

In this appendix, we show that if there is an extracellular shunt, the
measured impedance may be resistive, even with a non-resistive medium.

\begin{figure}[bht!]
 \centering
 \includegraphics[width=0.7\columnwidth]{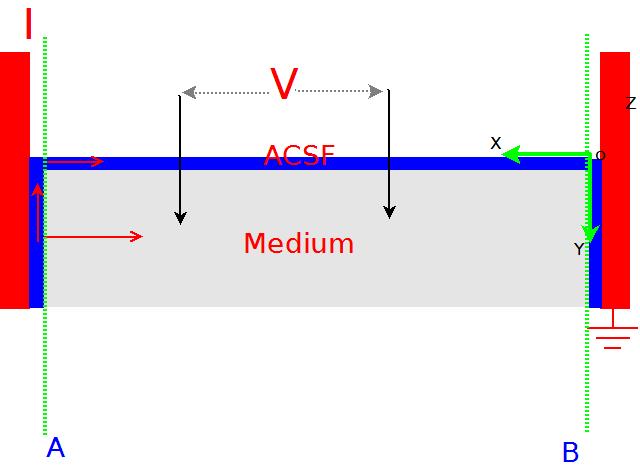}

 \caption{Scheme of an experimental setup for measuring impedances in
   biological tissue.  The injecting electrode (left) and reference
   electrode (right) are in contact with both the biological medium
   (gray) and the liquid layer (ACSF, blue), and thus the current flows in
   both (red arrows).  In a 4-electrode setup, two passive electrodes
   (middle, black arrows) are inserted in between the injecting and
   reference electrodes and are used to measure the voltage
   difference. }

 \label{imp4}
\end{figure}

We start from the scheme in Fig.~\ref{imp4}, where the injecting
electrode (left) and the reference electrode (right) are in contact
with the extracellular medium, as well as with the surrounding CSF.
For modeling such a ``macroscopic'' measurement, one must use a
mean-field version of Maxwell equations~\cite{BedDes2011a,BedDesMean}.  According
to the quasi-static electric approximation in mean-field, the electric
potential is solution of the Laplace equation: 
\b{\begin{equation}
    \nabla^2 V =0
\label{eq1}
\end{equation}} 
when the size of the volume element considered for the mean-field is
sufficiently large.  Note that this equation applies to the
extracellular medium, as well as to the surrounding CSF, and its
solution is unique for given boundary conditions. Also note that this
equation is the same in time or Fourier frequency space, because the
Laplacian is a linear and time independent operator.  In the
following, we use this equation in Fourier frequency space for more
convenience.

We now consider a typical setup to measure the impedance (or
admittance) of biological tissue, as schematized in Fig.~\ref{imp4}.
Our assumption is that the electrode is surrounded by a liquid layer of
CSF (schematized in blue), and that part of the current flows through
the medium, but also through this layer (red arrows).  Assuming that
the thickness of the tissue (slice or {\it in vivo}) is large compared
to the spatial scale of the mean-field, the electric potential must be
solution of the 2D Laplace equation: \b{
\begin{equation}
  \frac{\partial^2 V}{\partial x^2 } + \frac{\partial^2 V}{\partial y^2 } =0 ~ .
  \label{eq2}
  \end{equation}}

At a very short distance of the current injecting electrodes, we have 
the following constraint (by symmetry): 
\b{
\begin{equation}
\frac{\partial V(l_x,y)}{\partial x}
=
\frac{\partial V(0,y)}{\partial x}
\label{eq3}
\end{equation}}
when the electrodes are identical.

We now solve this system for the region between plane $A$ and $B$ (see
Fig.~\ref{imp4}).  According to the Stone-Weierstrass theorem, the
general solution of the 2D Laplace equation on a compact domain can be
represented by a two-variable series with integer exponents.  One can
group the terms of similar degree to form homogeneous polynomials and
calculate their coefficients so that they are solution of Laplace
equation.

 \b{
 \begin{equation}
 \left \{
 \begin{array}{rcl}
 P_0 &=&1 \\
 P_1 &=&a_1x+ b_1y\\
 P_2  &=&a_2(x^2-y^2) +b_2xy \\
 P_3 &=& a_3(x^3+3xy^2)+b_3(y^3+3x^2y) \\
 P_{\cdots} &=& \cdots 
 \end{array}
 \right .
 \end{equation}}

This method is equivalent to the construction of the particular
solutions of 3D Laplace equation using spherical
polynomials~\cite{smirnov}. Thus, we can write the solution of Laplace
equation (\ref{eq2}) as: \b{
\begin{equation} 
V(x,y) = \sum_{i=0}^{\infty }C_iP_i(x,y) ~ .
\label{eq5}
\end{equation}}

The symmetry condition (Eq.~\ref{eq3}) implies that \b{$C_i =0$} for
\b{$i>2$} and \b{$a_2=0$}, so that the general solution is given by:
  \b{
  $$
  \begin{array}{rcl}
  V(x,y;\omega) & = & C_1a_1x +C_1b_1y + C_2b_2(xy) \\
  & = & D_1x +D_2y + D_3xy
  \end{array} ~ .
  $$}

Note that here, the origin is placed on the reference electrode (see
Fig.~\ref{imp4}), which implies \b{$V(0,0)=0$}, so that we necessarily
have \b{$C_o=0$} in the expression (\ref{eq5}). It follows that the 
electric field is given by:
\b{
\begin{equation}
\vec{E} = -\nabla V = -(D_1 + D_3y)\hat{e}_x -(D_2+D_3x)\hat{e}_y ~ ,
\label{eq6}
\end{equation}}
and the respective current densities in CSF and in the medium are
given by:
\b{
\begin{equation}
\left \{
\begin{array}{rcl}
\vec{j}_{CSF}^g & = & -\gamma_{CSF}[(D_1 + D_3y)\hat{e}_x +(D_2+D_3x)\hat{e}_y] \\
\vec{j}_{medium}^g & = & -\gamma_{medium}[(D_1 + D_3y)\hat{e}_x +(D_2+D_3x)\hat{e}_y ]
\end{array}
\right .
\label{eq7}
\end{equation}}
We can derive the following expression for the current:
\b{
\begin{equation}
\left \{
\begin{array}{rcl}
I_{CSF}^g & = & \gamma_{CSF}(D_1 + \frac{1}{2}D_3l_y^{CSF})A_{CSF} \\
I_{medium}^g & = & \gamma_{medium}(D_1 + \frac{1}{2}D_3l_y^{medium})A_{medium} 
\end{array}
\right . ~ ,
\label{eq8}
\end{equation}}
where \b{$A_{CSF}$} is the area of the CSF layer along the $YZ$
plane, and \b{$A_{medium}$} is the area of the extracellular medium
(along the $YZ$ plane as well).  \b{$l_y^{CSF}$} is the thickness
of the CSF layer, and \b{$l_y^{medium}$} is the thickness of the
medium layer.

To keep the formalism as general as possible, and allow the impedance
of the extracellular medium to be non-resistive, we use the
generalized current conservation law.  Applying this current
conservation implies that \b{$I_{CSF}^g +I_{medium}^g = I^g$} does
not depend on $x$.

We can now evaluate \b{$D_1$} from the potential difference between
position (0,0) and ($l_x$,0) (see Fig.~\ref{imp4}). We have
\b{
\begin{equation}
\Delta V_A^B(y=0) = -\int_{(0,0)}^{(l_x,0)} \vec{E}\cdot d\vec{s}=D_1l_x ~ .
\label{eq9}
\end{equation}} 
Similarly, we can evaluate \b{$D_3$}, the potential difference between 
$(0,l_y)$ and $(l_x,l_y)$. We have
\b{
\begin{equation}
\Delta V_A^B(y=l_y) = -\int_{(0,l_y)}^{(l_x,l_y)} \vec{E}\cdot d\vec{s}=D_1l_x +D_3l_yl_x ~ .
\label{eq10}
\end{equation}}
Thus, we can write
\b{
\begin{equation}
\left \{
\begin{array}{ccccc}
D_1 &=& \frac{\Delta V_A^B(y=0)}{l_x}  \\\\\\
D_3 &= & \frac{\Delta V_A^B(y=l_y)-\Delta V_A^B(y=0)}{l_yl_x} 
\end{array}
\right .
\label{eq11}
\end{equation}
}
By assuming 
\b{$$<\Delta V_A^B> =\frac{1}{2}[\Delta V_A^B(y=0) +\Delta V_A^B(y=l_y)]  $$} 
 we can write (see Eq.~\ref{eq8}):
\b{
\begin{equation}
\begin{array}{rcl}
I^g & = & Y_{eq}\Delta V_A^B(y=0) \\ &=& (Y_{CSF}+Y_{medium})\Delta V_A^B(y=0) \\
    & = & (\gamma_{CSF}\frac{A_{CSF}}{l_x} +\gamma_{medium}\frac{A_{medium}}{l_x})\Delta V_A^B(y=0)
\end{array} ~ ,
\label{eq12}
\end{equation} }
where we assume that \b{$<\Delta V_A^B>~\approx ~\Delta V_A^B(y=0)$},
which is equivalent to neglect the electrode impedance.  Note that
neglecting the electrode impedance augments the ratio of current that
goes through the medium, compared to the current that flows through
the CSF, and thus this approximation diminishes the shunting effect.
\b{$Y_{CSF}$} and \b{$Y_{medium}$} are the macroscopic admittances of
the CSF and extracellular medium, respectively.  Note that the
``microscopic'' admittance is usually called \b{$\gamma $}, while
\b{$Y$} is the macroscopic admittance, as usually defined in
electronics for example.  Once the current is fixed, for example by a
current source, the knowledge of \b{$Y_{eq}(\omega )$} gives access to
\b{$<\Delta V_A^B>$} et , which allows one to determine \b{$V(x)$} at
every point in space.  Thus, the measure of the global current and the
potential difference between planes $A$ and $B$ does not give
information about each macroscopic admittance, but only about a global
admittance (the sum of each admittance in the system).

We now examine different possible scenarios for the respective values
of these admittances.

\textbf{Scenario 1: small medium admittance}.  If we assume that the admittance
of the medium is much smaller than that of CSF, we have:
\b{
$$
I \approx Y_{CSF} \Delta V_A^B ~ .
$$}

\textbf{Scenario 2: large medium admittance}.  If the medium admittance is larger
than that of CSF, we have:
\b{
$$
I \approx Y_{medium} \Delta V_A^B  ~ .
$$}

\textbf{Scenario 3: admittances of comparable magnitude.}  If we have nearly equal 
admittances, then we have:
\b{
$$
\begin{array}{rcl}
I & = & (Y_{medium}+Y_{CSF}) \Delta V_A^B \\
 & \approx & 2Y_{medium}\Delta V_A^B = 2Y_{CSF}\Delta V_A^B
\end{array}  ~ .
$$}

Thus, we see that the measured global admittance highly depends on the
relative admittance of the medium and CSF.  For example, finding a
weak frequency dependence of the measurement (as in \cite{Logothetis})
may mean that the medium is resistive, but it could also mean that
\b{$Y{medium}>> Y_{CSF}$} (Scenario 1).  Recent measurements
\cite{gomes2016} suggest that indeed \b{$Y{medium}$} is very high, and
the experiments reported here (Fig....) suggest that a significant
shunt is present with metal electrodes, so \b{$Y{CSF}$} is likely to
be small.  We expect that this admittance will be small for large
electrodes and will be larger for small electrodes such as
micropipettes.

In a four-electrode measurement setup, from Eq.~\ref{eq2}, we have:
\b{
\begin{equation}
V(x) = \frac{I}{Y_{CSF} +Y_{medium}}\frac{x}{l_x}
\end{equation}}
such that the voltage difference measured by the two central electrode
is given by:
 \b{
\begin{equation}
\Delta V = \frac{I}{Y_{CSF} +Y_{medium}}\frac{\Delta x}{l_x} ~ ,
\end{equation} }
where \b{$\Delta x$} is the distance between the two central electrodes.

We see that with a 4-electrode setup, the measured impedance will also
be dependent on the relative values of the admittance of the medium
and CSF.  Thus, even in such a setup, if a significant fraction of
the current flows through the CSF, the impedance of the medium may be
invisible in practice.  In the main text, we suggest a method to
prevent this possible source of error in the measurement.




\label{Bibliographie-Debut}

\end{document}